\documentclass[conference]{IEEEtran}
\IEEEoverridecommandlockouts
\usepackage{cite}
\usepackage{amsmath,amssymb,amsfonts}
\usepackage{algorithmic}
\usepackage{graphicx}
\usepackage{textcomp}
\usepackage{xcolor}
\def\BibTeX{{\rm B\kern-.05em{\sc i\kern-.025em b}\kern-.08em
    T\kern-.1667em\lower.7ex\hbox{E}\kern-.125emX}}
    
\usepackage{multirow}
\usepackage{ctable}
\usepackage[hidelinks]{hyperref}

\begin{document}

\title{Assessing Fatigue with Multimodal Wearable Sensors and Machine Learning}


\author{\IEEEauthorblockN{Ashish Jaiswal, Mohammad Zaki Zadeh, Aref Hebri, and Fillia Makedon}
\IEEEauthorblockA{Department of Computer Science \& Engineering\\
University of Texas at Arlington\\
Email: ashish.jaiswal@mavs.uta.edu}
}

\maketitle

\begin{abstract}
Fatigue is a loss in cognitive or physical performance due to physiological factors such as insufficient sleep, long work hours, stress, and physical exertion. It adversely affects the human body and can slow reaction times, reduce attention, and limit short-term memory. Hence, there is a need to monitor a person's state to avoid extreme fatigue conditions that can result in physiological complications. However, tools to understand and assess fatigue are minimal. This paper primarily focuses on building an experimental setup that induces cognitive fatigue (CF) and physical fatigue (PF) through multiple cognitive and physical tasks while simultaneously recording physiological data. First, we built a prototype sensor suit embedded with numerous physiological sensors for easy use during data collection. Second, participants' self-reported visual analog scores (VAS) are reported after each task to confirm fatigue induction. Finally, an evaluation system is built that utilizes machine learning (ML) models to detect states of CF and PF from sensor data, thus providing an objective measure. Our methods beat state-of-the-art approaches, where Random Forest performs the best in detecting PF with an accuracy of 80.5\% while correctly predicting the true PF condition 88\% of the time. On the other hand, the long short-term memory (LSTM) recurrent neural network produces the best results in detecting CF in the subjects (with 84.1\% accuracy, 0.9 recall).
\end{abstract}



\section{INTRODUCTION}

Fatigue is a state of weariness that develops over time and reduces an individual's energy, motivation, and concentration. Fatigue can be classified into three types: Acute fatigue is caused by excessive physical or mental exertion and is alleviated by rest. Normative fatigue is influenced by changes in circadian rhythm and daily activities. In contrast, Chronic fatigue is primarily caused by stress or tension on the body and is less likely to be relieved by rest alone. While a variety of factors influence human fatigue in the real world, factors affecting sleep and the circadian system have a high potential to contribute to fatigue \cite{ji2006probabilistic}.

Severe or chronic fatigue is usually a symptom of a disease rather than the result of daily activities. Some diseases, such as Multiple Sclerosis (MS) \cite{krupp1988fatigue}, Traumatic Brain Injury (TBI) \cite{belmont2006fatigue}, and Parkinson's Disease (PD) \cite{hagell2009towards}, have fatigue as a major symptom. Physical and cognitive fatigue are the two types of fatigue. Physical fatigue (PF) is most commonly caused by excessive physical activity and is usually associated with a muscle group or a general feeling of fatigue in the body \cite{chaudhuri2004fatigue}. Cognitive fatigue (CF), on the other hand, can occur as a result of intense mental activity, resulting in a loss of cognition with decreased attention and high-level information processing \cite{meier2014developmental}. In the real world, however, there is no clear distinction between what causes both types of fatigue. Workers with heavy machinery, for example, may require both cognitive skills and physical labor to complete a task that may induce both PF and CF at the same time.

Researchers have previously attempted to assess both types of fatigue separately by approaching them differently. One of the most common methods of studying fatigue is to analyze the participants' subjective experience by having them fill out surveys rating their current state of fatigue. Although these methods have successfully quantified human fatigue, they are frequently prone to human bias and poor data collection methods. An aviation study \cite{bendak2020fatigue} discovered that 70-80 percent of pilots misrepresented their fatigue level. As a result, relying solely on a subjective measure from the participants may raise safety concerns. It is where physiological sensors, which provide objective measures of fatigue, come into play. To study fatigue, data collected from sensors such as electrocardiograms (ECG) \cite{huang2018detection}, electroencephalograms (EEG)\cite{jap2009using}, electrodermal activity/galvanic skin response (EDA/GSR) \cite{dawson2011skin}, and electromyograms (EMG) \cite{cifrek2009surface} have been commonly used.

\begin{figure*}[ht]
    \centering
    \includegraphics[width=0.9\linewidth,height=40mm]{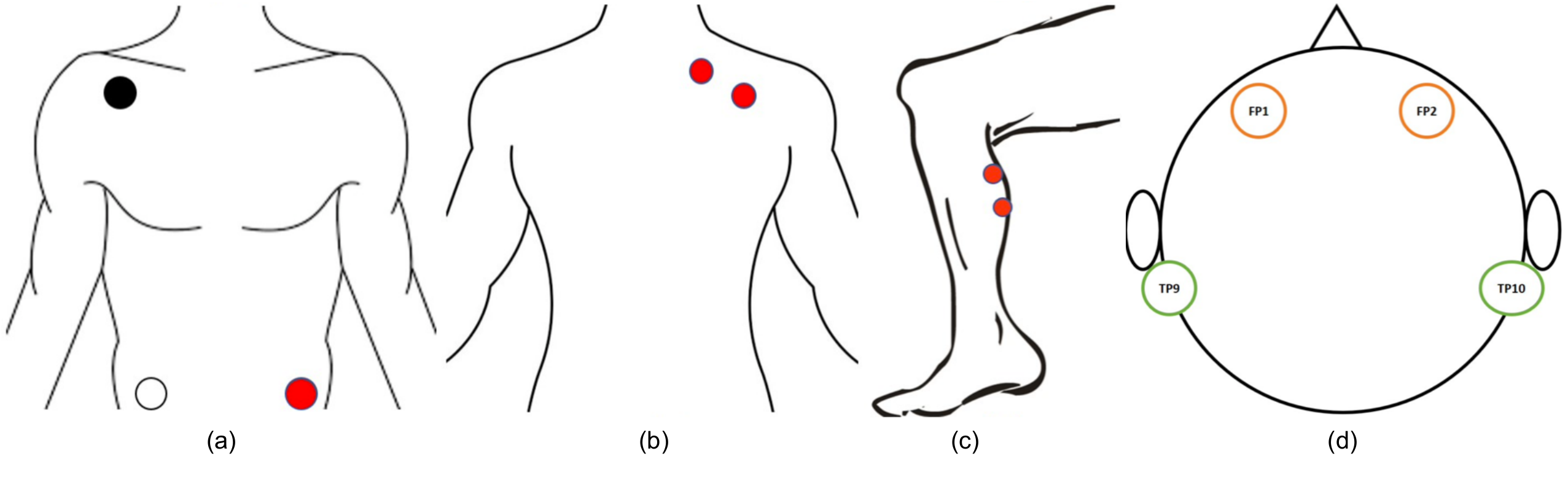} 
    \caption{Sensor placements on the human body (a) \textbf{ECG}: right shoulder, the left and the right hip forming Einthoven’s triangle \cite{einthoven1950direction} (b) \textbf{EDA/GSR} electrodes on the left shoulder to record the skin conductivity, (c) \textbf{EMG} electrodes recording muscle twitches from the right calf, (d) \textbf{EEG} sensor positions in the 10-10 electrode system used by MUSE. It records data from the TP9, AF7, AF8, and TP10 positions in the system.}
    \label{fig:sensors_placement_combined}
\end{figure*}

In this paper, we created an experimental setup that includes multiple standardized cognitive tasks (N-Back tasks \cite{hopstaken2015multifaceted,bailey2007relationship,guastello2015catastrophe}) and a physical task (running on a treadmill \cite{garcia2014effects,myles1985sleep}) to induce both CF and PF simultaneously. When the participants completed each of the assigned tasks, data modalities such as ECG (for heart rate variability), EMG (for muscle activation), EDA/GSR (for skin conductivity and emotional arousal), and EEG (for electrical brain signals) were recorded. We combined multiple physiological data modalities to assess both CF \& PF and correlate objective measures from the sensors with the participants' self-reported subjective visual analog scores (VAS). The self-reported scores were analyzed to verify the induction of fatigue in the participants. Finally, several statistical features were extracted from the collected signals, and machine learning (ML) models were trained to predict the participant's fatigue levels. The major contributions of the paper are highlighted as follows:

\begin{itemize}
    \item A novel experimental setup that integrates standardized cognitive and physical tasks to induce fatigue and record multivariate physiological sensor signals to analyze the state of fatigue in a subject
    \item Ablation study to understand what combination of sensors produces the best results in detecting both cognitive and physical fatigue
    \item Statistical feature extraction and machine learning models to classify the state of fatigue from sensor data using different window-sizes technique
    \item A well-accumulated dataset consisting of EEG, EDA/GSR, EMG, and ECG signals from 32 healthy subjects with their self-reported fatigue states. To access the dataset, visit \textcolor{blue}{\href{https://drive.google.com/file/d/1KXbK0tQZ46-jF3SHP13MkRbyOYKA-Tdz/view?usp=sharing}{[Dataset Link]}}.
\end{itemize}

The rest of the paper is structured as follows: We first present the related work in section \ref{section:related_work}. Section \ref{section:data_collection} explains the data collection process followed by data processing techniques for different sensor data in section \ref{section:data_processing}. We then describe the experiments and techniques we inherited in our project along with the obtained results in section \ref{section:results}. Finally, we conclude the paper and propose future work in section \ref{section:conclusion}.

\section{RELATED WORK}
\label{section:related_work}

Wearable sensors have been studied for their use in fatigue assessment in occupational settings, healthcare, sports, and exercise applications. The current literature emphasizes using wearable systems to collect physiological data from subjects, laying the groundwork for fatigue quantification using behavioral and physiological signal-based methods.

In recent years, studies have attempted to use wearable sensors and additional modalities to assess fatigue during daily activities \cite{babu2018multimodal, babu2018facial, ramesh2020multi}. For example, Aryal et al. \cite{Aryal2017monitor} used wearable sensors to record heart rate, skin temperature, and EEG to assess fatigue in construction workers. In addition, Maman et al. \cite{Sedighi2017data} conducted a study that focused on detecting physical fatigue by measuring heart rate and analyzing body posture during various physical tasks. Similarly, in \cite{al2016hrv}, the authors presented a new operator-fatigue analysis and classification method based on heart rate variability (HRV) using low-cost wearable devices that identified the operator's alertness/fatigue states. Finally, not being limited to constrained experimental setups, the authors in \cite{bai2020fatigue} collected ECG and Actigraphy data from free-living environments to assess objective fatigue based on self-reported scores.


Although most studies have focused solely on either CF or PF, some researchers have attempted to study both types of fatigue concurrently. For example, Marcora et al. \cite{Marcora2009Mental} discovered that mental fatigue affects exercise tolerance, while Xu et al. \cite{Xu2018How} proposed that physical activities affect mental fatigue. Similarly, a single wearable sensor that measures acceleration and ECG during a self-paced trial run was used to assess CF and PF by \cite{russell2021predicting}. In addition to wearable sensors, work has been done to assess CF from visual data such as fMRI scans \cite{zadeh2020towards, jaiswal2021understanding}.


A review of fatigue sensors \cite{adao2021fatigue} found out that motion (MOT), electroencephalogram (EEG), photoplethysmogram (PPG), electrocardiogram (ECG), galvanic skin response (GSR), electromyogram (EMG), skin temperature (Tsk), eye movement (EYE), and respiratory (RES) sensors were the most effective. Among the proposed fatigue quantification approaches, supervised machine learning models were found to be dominant. This paper expands on previous research to look into different wearable sensors for fatigue assessment. While previous research has primarily focused on assessing CF or PF separately, our research combines multiple physiological sensors to assess CF and PF at a larger scale.

\begin{figure}[ht]
    \centering
    \includegraphics[width=1.0\linewidth]{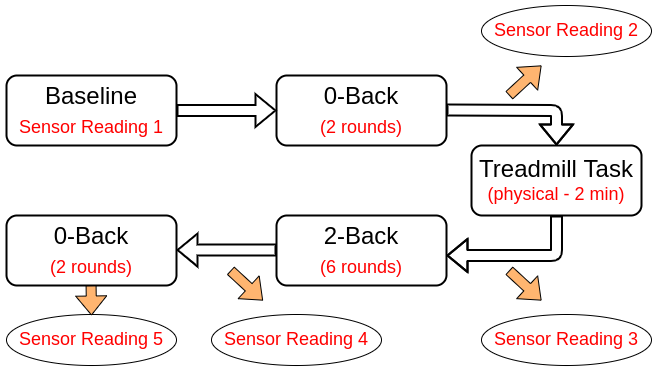}
    \caption{Flow diagram of the tasks performed by a participant.}
    \label{fig:task_flow_diagram}
\end{figure}

\begin{figure*}[ht]
    \centering
    \includegraphics[width=1.0\linewidth,height=60mm]{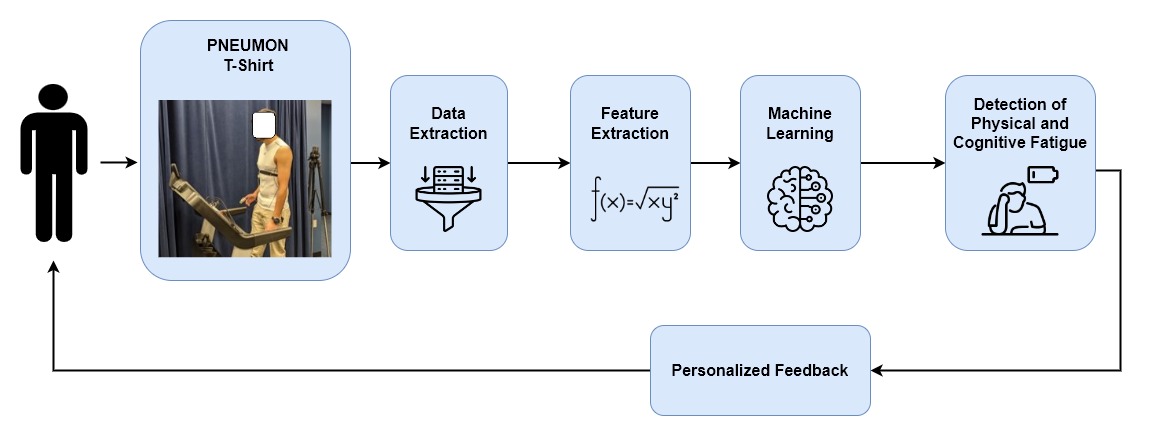}
    \caption{System Flow Diagram: Data collection using the sensors attached on the PNEUMON t-shirt and MUSE S worn by one of the participants while performing tasks presented in Fig. \ref{fig:task_flow_diagram}. Features extracted from the recorded signals were used to train ML models for detection of state of CF and PF.}
    \label{fig:system_flow_diagram}
\end{figure*}

\section{DATA COLLECTION} \label{section:data_collection}

We built an experimental setup around a custom-built wearable t-shirt (Pneumon) (presented in Fig. \ref{fig:va_tshirt}) that was used to record physiological data using the attached sensors and a MUSE S headband \cite{muses} (as shown in Fig. \ref{fig:muse_electrodes}). In Fig. 3, although additional sensors are connected to the shirt, such as Microphone, Breathing Band, and Oximeter, we ignored those signals for fatigue detection as they added negligible value to our analysis in this paper. The suit is made using a stretchable Under Armour shirt \cite{underarmour}. The advantage of using a shirt with embedded sensors is its ease of use during data collection and practical application during day-to-day use. In addition, the sensors are hidden behind a detachable covering, so they can easily be removed while washing the shirt. Data from 32 healthy people (18-33 years old, average age 24 years, 28/72\% female/male) were collected in two separate study sessions. In addition, brain EEG data was collected using a MUSE S headband sensor. Participants were required to wear the t-shirt and the MUSE headband throughout the experiment.

\begin{figure}[ht]
    \centering
    \includegraphics[width=1.0\linewidth]{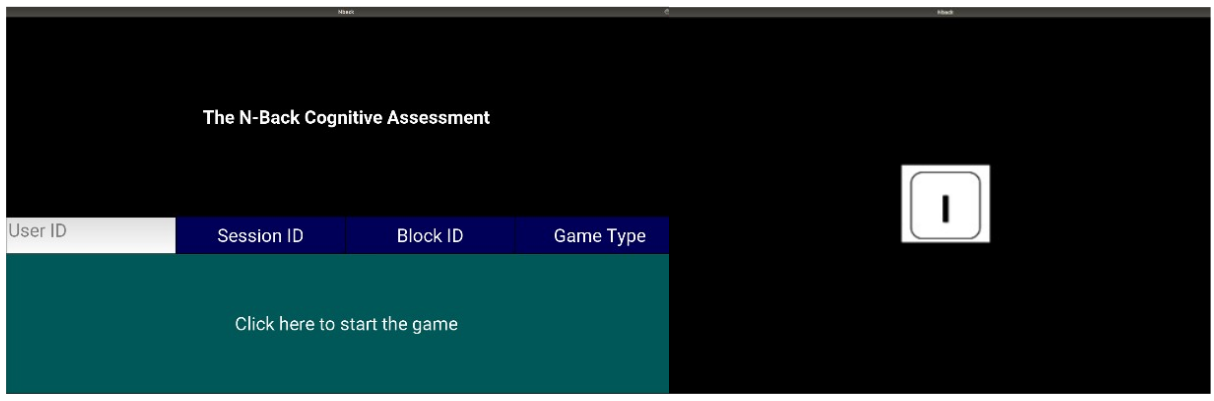}
    \caption{Graphical User Interface built for the N-Back tasks with an example image of a letter during a game round on the right.}
    \label{fig:n_back_GUI}
\end{figure}

\begin{figure}[ht]
    \centering
    \includegraphics[width=1.0\linewidth]{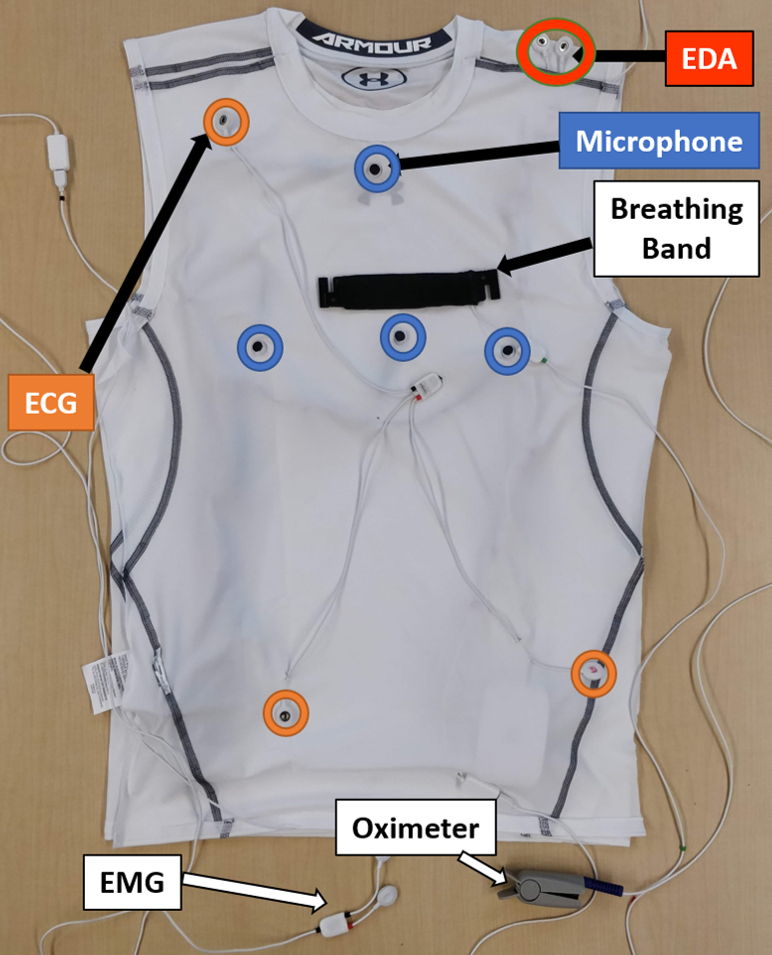}
    \caption{A prototype of the sensor shirt (inside-out view): Physiological sensors are embedded in the shirt that remain in contact with the subject's torso during data collection using an adhesive tape. Multiple sensory signals are collected using different combinations of the sensors present. \textbf{In this paper, we ignore the microphone, oximeter, and the breathing band signals.}}
    \label{fig:va_tshirt}
\end{figure}

The researchers began by taking a baseline reading from the sensors while the subject stood still for one minute. The experiment's goal was to induce CF and PF while collecting sensory data throughout the process. Next, the participants were asked to perform multiple sets of N-Back tasks to induce CF, as shown in Fig. \ref{fig:task_flow_diagram}. The GUI of the N-back game as observed by the participant is shown in Fig. \ref{fig:n_back_GUI}. The subject is shown a series of letters one after the other in these tasks. The subject's goal is to determine whether the current letter matches the letter presented N steps back. If it does, the subject must carry out the action specified (pressing the space bar on the keyboard). To induce PF in the subjects, they were asked to run for 2 minutes on a treadmill (speed - 5mph, incline - 10\%). Additional data was collected (sensor reading 3 in Fig. \ref{fig:task_flow_diagram}) after the subjects stood still for 90 seconds following the completion of the physical task.

\begin{figure}[ht]
    \centering
    \includegraphics[width=0.7\linewidth,scale=0.7]{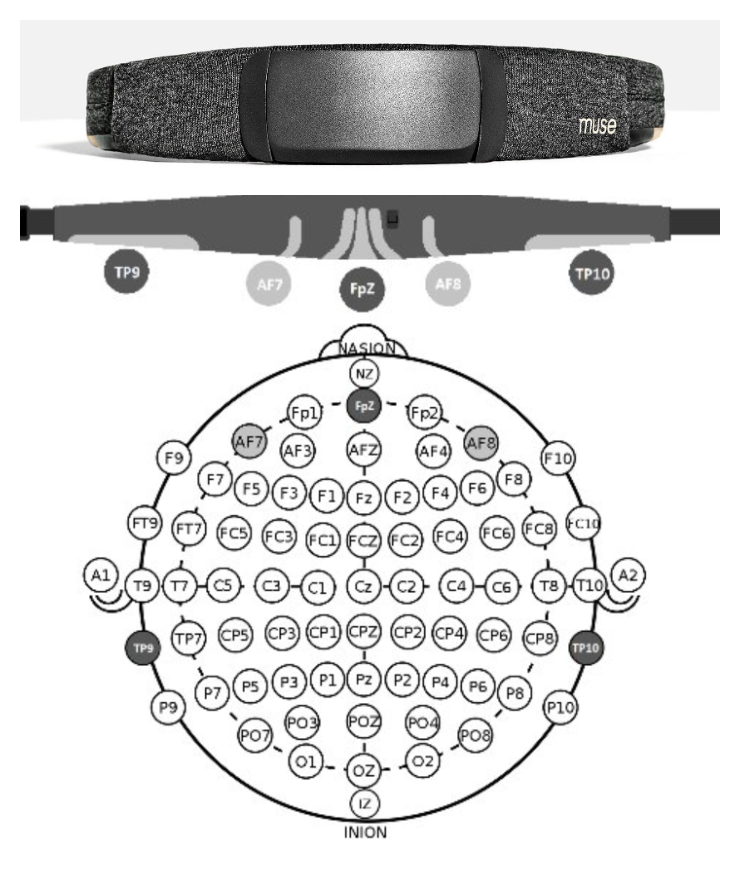}
    \caption{Electrode placement comparison between MUSE and the international 10-20 system. Top: Commercial MUSE S Headband. Bottom: 10-20 Electrode Placement System}
    \label{fig:muse_electrodes}
\end{figure}

The study was divided into two sessions on separate days for each participant. Subjects were asked to come in the morning for one session and in the evening for another. It eliminated the effect on the data caused by the time of the day. The tasks performed in both sessions were identical, with the only difference being the order they were completed. The first session followed the flow depicted in Fig. \ref{fig:task_flow_diagram} whereas the second session prioritized the cognitively challenging 2-Back game before the physical task. It dismisses PF's reliance on CF and allows us to collect more robust data for analysis. Each round of 0-back and 2-back tasks lasted between 80.4 seconds and  144.3 seconds respectively (on average) during the data collection process.


Participants were asked to complete a brief survey indicating their current physical and cognitive fatigue levels following each task. In addition, they reported visual analog scale (VAS) scores ranging from 1 to 10 for the following questions:

\begin{enumerate}
    \item Describe your overall tiredness on a scale of 1-10.
    \item How physically fatigued do you feel on a scale of 1-10?
    \item How cognitively fatigued do you feel on a scale of 1-10?
    \item How sleepy or drowsy do you feel on a scale of 1-10?
\end{enumerate}

Fig. \ref{fig:survey_plots} depicts the distribution of self-reported VAS scores after each task. We divided the scores from 1 to 10 into three categories for simplification while plotting: None ($<$4), Moderate ($\geq4$ and $=$7), and Extreme ($>$7). The majority of the participants gradually began to feel cognitively fatigued after completing each task. In fact, after the fourth block, CF appears to be induced in more than 80\% of the subjects (Fig. \ref{fig:survey_plots}(d)), supporting the hypothesis on which we based our experimental setup. Similarly, as shown in Fig. \ref{fig:survey_plots}(c), the physical task was able to induce at least moderate PF in more than 90\% of the subjects (c). Thus, data collected during the fourth and fifth blocks of the study were considered a state for CF in the participants, whereas data collected immediately following the physical task was considered a PF condition. There may be some biases in the subjective scores collected. However, the majority of the scores recorded from the participants (as shown in Fig. \ref{fig:survey_plots}) seem to support the hypothesis on which the whole experiment was based.

\section{DATA PROCESSING}
\label{section:data_processing}

\subsection{EEG}

\begin{figure}[ht]
    \centering
    \includegraphics[width=1.0\linewidth]{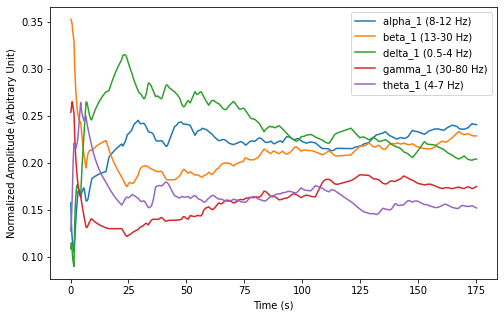}
    \caption{Raw Amplitude plot of Frequency bands extracted from electrode at AF7 position (on MUSE) from a sample raw EEG signal from one of the subjects. The readings were collected during one of the 2-Back tasks undergoing for a little under 3 minutes.}
    \label{fig:eeg_bands}
\end{figure}

We used MUSE S headset to collect EEG signals while the subjects performed different tasks during the experiment. It consists of 4 electrodes (AF7, AF8, TP9, TP10) that are in contact with different regions of the head, as shown in Fig. \ref{fig:sensors_placement_combined}(d). EEG signals quantify electrical activity in the brain. These signals are generally decomposed into five different frequency bands: alpha, beta, delta, gamma, and theta, as shown in Fig. \ref{fig:eeg_bands} where each band signifies a different state of the brain. For example, delta waves occur in the frequency range of 0.5 Hz to 4 Hz and are present during deep sleep, while beta waves occur between 13 Hz to 30 Hz and are associated with active thinking. Similarly, other waves are associated with -- alpha waves (8-12 Hz): normal awake conditions, gamma waves (30-80 Hz): sensory perception integration, and theta waves (4-7 Hz): drowsiness and early stages of sleep. 50-60 Hz frequencies were processed beforehand to avoid power line interference on the signals.

\begin{figure*}[ht]
    \centering
    \includegraphics[width=1.0\linewidth]{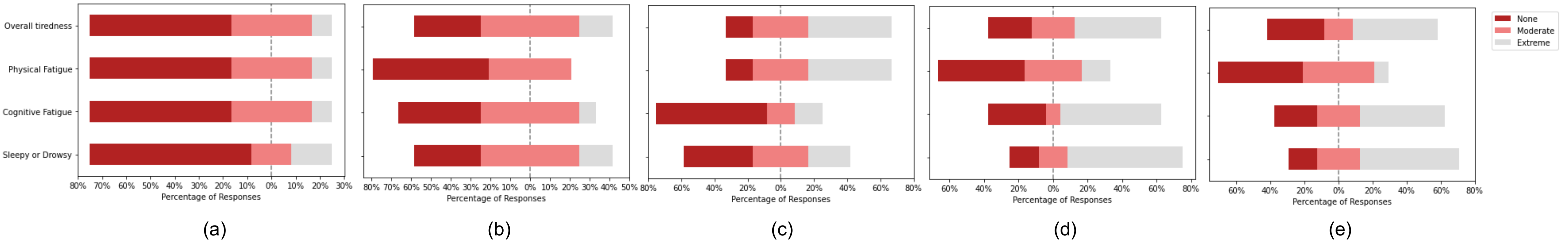} 
    \caption{Likert plots for survey responses from subjects after each block of tasks (before taking the sensor readings) as shown in Fig. \ref{fig:task_flow_diagram}. (a) Initial survey at the start of the session showing that less to none fatigue was found in most of the participants. (b) Response after the first 0-Back task and before sensor reading 2 showing slight increase in CF for some participants. (c) Response after the treadmill task verifying that PF is induced in most of the participants. (d) Response after the first 2-Back task indicating prominent increase in CF (e) Shows that CF continuing to be persistent even during the easier 0-Back task}
    \label{fig:survey_plots}
\end{figure*}

The different EEG bands and the raw signals were streamed by the MUSE SDK using a UDP server and were recorded from all four electrodes in a csv file. We calculated statistics such as mean, standard deviation, min, max, and median using the sliding window technique for each band from the respective electrodes for feature extraction. 

\subsection{ECG, EDA, and EMG}

Physiological sensors were used in the wearable t-shirt, as shown in Fig. \ref{fig:system_flow_diagram} to collect ECG, EDA, and EMG signals simultaneously during the entire experiment. Fatigue can harm the cardiovascular system, endocrine system, and brain. Therefore, these multi-modal signals help keep track of the subject's physical state and can provide quality information on whether the person is feeling fatigued or not.

ECG signals show changes in the cardiovascular system by reflecting the heart's electrical activity. It contains essential information about the cardiac pathologies affecting the heart, characterized by five peaks known as fiducial points, which are represented by the letters P, Q, R, S, and T \cite{richley2013new}. Studies have shown that fatigue affects the body's cardiovascular response \cite{nelesen2008relationship}. We used Einthoven's triangle approach to record the ECG signals \cite{einthoven1950direction}. Eithoven's triangle is an imaginary formation of three limb leads in a triangle used in electrocardiography, formed by the two shoulders and the pubis. It can be helpful in the identification in incorrect placement of the sensor leads.

On the other hand, EDA (often referred to as galvanic skin response, or GSR) reflects the activity of the sympathetic nervous system, which is dependent on physiological and emotional activation by measuring the skin conductivity of the body. EDA signals can also be reflective of the intensity of our emotional state. Therefore, measuring this signal allows us to identify psychological or emotional arousal episodes. Finally, EMG signals record the change in voltage between two electrodes during muscle contraction and relaxation. PF has been shown to affect muscle activity and can lead to reduced muscle activation \cite{rota2014influence}.

To remove unwanted noise from the ECG signals, we used Pan, and Tompkins QRS detection algorithm \cite{pan1985real}. First, we cleaned the signals using a high-pass Butterworth filter with a fixed cut-off frequency of 0.5 Hz. After this, a notch filter was applied to the cleaned signal to remove components with 50 Hz frequency and avoid power line interference. Next, RR intervals were extracted using the R\_Peaks from the signal and were cleaned by removing the outliers. The missing values were substituted using the linear interpolation method. Finally, 113 time-domain and frequency-domain features, including heart rate variability (HRV) metrics, were extracted to be used for training ML models.

EDA signals were first passed through a low-pass Butterworth signal with a cut-off frequency of 3 Hz. EDA signals can be decomposed into two components: phasic and tonic \cite{braithwaite2013guide}. Since the phasic component of the EDA signal is the faster-changing element of the signal based on physiological changes due to a stimulus, it was extracted for processing. Skin Conductance Response (SCR) peaks were extracted as features from the cleaned EDA signals. Similarly, time domain and frequency domain features were extracted from the EMG signals. Most of the feature extraction was carried out using the package named Neurokit2 \cite{neurokit} for all three signals.

\begin{table*}[t]
    \parbox{.5\linewidth}{
        \centering
        \caption{Detection of CF with EEG Features}
        \begin{tabular}{|c|c|c|c|c|c|}
             \hline
             \multirow{2}{2em}{Model} & \multicolumn{4}{|c|}{Accuracy (Window Size)} & \multirow{2}{2.2em}{Avg. Recall}\\
             \cline{2-5}
             & 5s & 10s & 20s & Full Block & \\
             \hline
             Log Reg. & 69.7\% & 72.6\% & 71.3\% & 62.3\% & 0.76 \\
             \hline
             SVM & 73.1\% & 73.3\% & 71.7\% & 69.4\% & 0.81 \\
             \hline
             RF & 72.3\% & \textbf{81.9\%} & 79.1\% & 76.3\% & \textbf{0.89} \\
             \hline
             LSTM & 69.8\% & 71.8\% & 73.8\% & \textbf{81.9\%} & 0.82 \\
             \hline
        \end{tabular}
        \label{table:I}
    }
    \hfill
    \parbox{.5\linewidth}{
        \centering
        \caption{Detection of CF with ECG + EDA + EMG Features}
        \begin{tabular}{|c|c|c|c|c|c|}
             \hline
             \multirow{2}{2em}{Model} & \multicolumn{4}{|c|}{Accuracy (Window Size)} & \multirow{2}{2.2em}{Avg. Recall}\\
             \cline{2-5}
             & 5s & 10s & 20s & Full Block & \\
             \hline
             Log Reg. & 69.8\% & 70.1\% & 67.2\% & 65.3\% & 0.69 \\
             \hline
             SVM & 71.2\% & 71.7\% & 70.8\% & 70.1\% & \textbf{0.73} \\
             \hline
             RF & 74.8\% & \textbf{76.3\%} & 72.1\% & 70.9\% & 0.71 \\
             \hline
             LSTM & 62.2\% & 63.7\% & 68.9\% & 70.1\% & 0.69 \\
             \hline
        \end{tabular}
        \label{table:II}
     }
\end{table*}

\begin{table*}[t]
    \parbox{.5\linewidth}{
        \centering
        \caption{Detection of PF with ECG + EDA + EMG Features}
        \begin{tabular}{|c|c|c|c|c|c|}
             \hline
             \multirow{2}{2em}{Model} & \multicolumn{4}{|c|}{Accuracy (Window Size)} & \multirow{2}{2.2em}{Avg. Recall}\\
             \cline{2-5}
             & 5s & 10s & 20s & Full Block & \\
             \hline
             Log Reg. & 72.2\% & 72.2\% & 68.2\% & 62.9\% & 0.74 \\
             \hline
             SVM & 76.1\% & 79.6\% & 75.2\% & 73.1\% & 0.86 \\
             \hline
             RF & 79.9\% & \textbf{80.5\%} & 77.6\% & 77.2\% & \textbf{0.88} \\
             \hline
             LSTM & 64.2\% & 64.8\% & 62.7\% & 68.9\% & 0.79 \\
             \hline
        \end{tabular}
        \label{table:III}
    }
    \hfill
    \parbox{.5\linewidth}{
        \centering
        \caption{Detection of CF with EEG + ECG + EDA + EMG Features}
        \begin{tabular}{|c|c|c|c|c|c|}
             \hline
             \multirow{2}{2em}{Model} & \multicolumn{4}{|c|}{Accuracy (Window Size)} & \multirow{2}{2.2em}{Avg. Recall}\\
             \cline{2-5}
             & 5s & 10s & 20s & Full Block & \\
             \hline
             Log Reg. & 64.0\% & 66.9\% & 66.1\% & 60.4\% & 0.69 \\
             \hline
             SVM & 70.3\% & 74.6\% & 74.5\% & 70.3\% & 0.79 \\
             \hline
             RF & 67.9\% & 77.2\% & 76.8\% & 74.5\% & 0.81 \\
             \hline
             LSTM & 71.3\% & 74.2\% & 74.8\% & \textbf{84.1\%} & \textbf{0.90} \\
             \hline
        \end{tabular}
        \label{table:IV}
    }
\end{table*}

\section{EXPERIMENTATION AND RESULTS}
\label{section:results}

We extracted 100 statistical features from EEG signals and 169 combined features from ECG, EDA, and EMG for training the ML models. Based on the responses accumulated from the participants (in Fig. \ref{fig:survey_plots}), data collected in sensor readings 1, 2, and 3 (before the 2-Back tasks in Fig. \ref{fig:task_flow_diagram}) were labeled as "No CF" condition. In contrast, the data recorded during the final two readings (4 and 5, i.e., after the 2-Back rounds) were labeled as "CF" conditions. Similarly, data recorded right after the physical task (sensor reading 3) was labeled as "PF" condition when the subjects stood still. Meanwhile, readings 4 and 5 were not considered for PF analysis.

For training, instead of processing the entire signal for a task as a single input, we split the time signal into multiple slices based on different window sizes (5 seconds, 10 seconds, and 20 seconds). Each signal slice was labeled the same as the original signal label, and features were extracted. It also increased the number of input data points to the ML models during training. However, we also tested the models with entire signal blocks as inputs. Similarly, during inference, the input signal was broken down into smaller slices based on the window size chosen during training, and each slice was classified as one of the classes by the model. Finally, the whole signal block was classified based on the higher count of the class among the classified slices. This technique makes the model more robust toward any noise or outliers in the signals, as noise contained in some of the slices may not contribute much to the final classification result.

The entire dataset was randomly split into train (70\%, 22 subjects), validation (15\%, 5 subjects), and test (15\%, 5 subjects) sets.
We used stratified sampling while splitting the dataset to prevent dealing with an unbalanced dataset. In addition, 5-fold cross-validation was performed for each of the models. Four different ML models: Logistic Regression (Log Reg.), Support Vector Machines (SVM), Random Forest (RF), and Long Short-Term Memory (LSTM) recurrent neural network were used in the analysis. We used multiple combinations of features extracted from the signals to predict fatigue.

Since EEG signals represent activity in the brain that is related to cognitive functions, initially, only features extracted from the EEG data were used to predict CF, as shown in table \ref{table:I}. Similarly, we used features extracted from the physiological sensors (ECG, EDA, and EMG) to train models that predict both CF and PF as presented in tables \ref{table:II} and \ref{table:III}. Finally, features from all the data modalities were combined and normalized for the detection of CF, represented by table \ref{table:IV}. For results in table \ref{table:IV}, we applied Principal Component Analysis (PCA) \cite{Ringnr2008WhatIP} to reduce dimensions to 189 features for the best results from the ML models.

\begin{table}[ht]
    \centering
    \caption{Comparison with SOTA Algorithms}
    \begin{tabular}{|c|c|c|c|c|}
        \hline
        \multirow{2}{2em}{\textbf{Fatigue Type}} & \multirow{2}{2em}{\textbf{Model}} & \multirow{2}{4em}{\textbf{Accuracy}} & \multirow{2}{4em}{\textbf{Avg. Recall}} & \multirow{2}{2em}{\textbf{Ref}}. \\
        & & & & \\
        \hline
        Physical & RF & $71.85$ & $0.72$ & \cite{luo2020assessment} \\
        \hline
        Physical & cCNN + RF & $71.40$ & $0.73$ & \cite{luo2020assessment} \\
        \hline
        \textbf{Physical} & \textbf{RF (Ours)} & $\mathbf{80.50}$ & $\mathbf{0.88}$ & Table \ref{table:III} \\
        \specialrule{.2em}{.1em}{.1em}
        Cognitive & RF & $64.69$ & $0.65$ & \cite{luo2020assessment} \\
        \hline
        Cognitive & RF & $66.20$ & $0.66$ & \cite{luo2020assessment} \\
        \hline
        \textbf{Cognitive} & \textbf{LSTM (Ours)} & $\mathbf{84.1}$ & $\mathbf{0.90}$ & Table \ref{table:IV} \\
        \hline
    \end{tabular}
    \label{table:V}
\end{table}

The first three models (Log Reg., SVM, and RF) were trained using the features extracted from the signals. However, an LSTM model (with 256 hidden layers) was trained on the raw signals directly as it can process time-series data. We used the same window-based approach to train the LSTM models, where the input size for the EEG signals was t x 20 x 1 (five frequency bands from each of the four electrodes). On the other hand, ECG, EDA, and EMG signals were combined to form t x 3 x 1 inputs. Finally, for all signals combined, the LSTM was trained on t x 23 x 1 inputs. Here, "t" represents the number of timesteps in the signal, which varies based on the window size. 

The average recall (Avg. Recall) presented in all four tables is the average recall for the "Fatigue" condition (either cognitive or physical) obtained across 5-fold cross-validation for each model. The best value obtained for each model among different window sizes was considered. With a CF prediction accuracy of 81.9\%, RF seems to perform the best using only the EEG features in table \ref{table:I}. We can see that it correctly predicts CF conditions 89\% of the time. The recall is an essential metric since our primary goal is to detect actual fatigue conditions in the subjects and avoid false negatives.

Similarly, RF seems to outperform the rest of the models when trained with features from ECG, EDA, and EMG to detect both CF and PF with respective accuracies of 76.3\% (recall=0.71) and 80.5\% (recall=0.88). Also, we can confirm that the window size of 10s seems to work the best for feature-based models, while the LSTM model performs the best when provided with the entire block of signal collected. Based on table \ref{table:IV}, LSTM performs the best when all four modalities are combined, and the whole signal block is provided to the model. Hence, for the detection of CF, feature engineering can be avoided entirely with four modalities combined and processed directly with an LSTM network.

Finally, there is only one study by Luo et. al \cite{luo2020assessment} that deals with fatigue detection in human subjects using wearable sensors. Their pilot study looked at multiple digital data sources on physical activity, vital signs, and other physiological parameters and their relationship to subject-reported non-pathological physical and mental fatigue in real-world settings. They demonstrated how multimodal digital data could be used to inform, quantify, and augment subjectively collected non-pathological fatigue measures. Based on the performance of their methodology with ours in Table \ref{table:V}, we can see that our methods outperform their approaches in detecting both physical and cognitive fatigue with models RandomForest (RF) and LSTM, respectively.

\section{CONCLUSION}
\label{section:conclusion}

This paper presents one of the preliminary works that utilizes a unique combination of physiological (ECG, EDA, EMG) and brain (EEG) sensors to detect CF and PF simultaneously. The flow of the tasks designed for data collection successfully induced CF ($>$ 80\% participants) and PF ($>$ 90\% participants) based on the reported subjective VAS from the participants. While the Random Forest classifier performed the best in detecting PF, the success of the LSTM model in predicting CF eliminates the need for extensive data pre-processing and feature extraction. Overall, the best models in the system failed to detect actual PF in less than 12\% of the cases (recall=0.88) and CF in less than 10\% (recall=0.90), with promising results. Further research directions will include visual sensor data to analyze facial expressions and gait movement that will aid in better prediction of fatigue. Additionally, including subjects with severe conditions impacted by fatigue can benefit the study in detecting symptoms related to those diseases.

\bibliographystyle{splncs03}
\bibliography{references}

\begin{thebibliography}{10}
\providecommand{\url}[1]{\texttt{#1}}
\providecommand{\urlprefix}{URL }

\bibitem{adao2021fatigue}
Ad{\~a}o~Martins, N.R., Annaheim, S., Spengler, C.M., Rossi, R.M.: Fatigue
  monitoring through wearables: a state-of-the-art review. Frontiers in
  physiology p. 2285 (2021)

\bibitem{al2016hrv}
Al-Libawy, H., Al-Ataby, A., Al-Nuaimy, W., Al-Taee, M.A.: Hrv-based operator
  fatigue analysis and classification using wearable sensors. In: 2016 13th
  International Multi-Conference on Systems, Signals \& Devices (SSD). pp.
  268--273. IEEE (2016)

\bibitem{underarmour}
Armour, U.: Men’s ua heatgear armour sleeveless compression shirt,
  \url{"https://www.underarmour.com/en-us/p/tops/mens\_ua\_heatgear\_armour\_sleeveless\_compression\_shirt/1257469.html"}

\bibitem{Aryal2017monitor}
Aryal, A., Ghahramani, A., Becerik-Gerber, B.: Monitoring fatigue in
  construction workers using physiological measurements. Automation in
  Construction  82,  154--165 (2017)

\bibitem{babu2018facial}
Babu, A.R., Cloud, J., Theofanidis, M., Makedon, F.: Facial expressions as a
  modality for fatigue detection in robot based rehabilitation. In: Proceedings
  of the 11th PErvasive Technologies Related to Assistive Environments
  Conference. pp. 112--113 (2018)

\bibitem{babu2018multimodal}
Babu, A.R., Rajavenkatanarayanan, A., Brady, J.R., Makedon, F.: Multimodal
  approach for cognitive task performance prediction from body postures, facial
  expressions and eeg signal. In: Proceedings of the Workshop on Modeling
  Cognitive Processes from Multimodal Data. pp. 1--7 (2018)

\bibitem{bai2020fatigue}
Bai, Y., Guan, Y., Ng, W.F.: Fatigue assessment using ecg and actigraphy
  sensors. In: Proceedings of the 2020 International Symposium on Wearable
  Computers. pp. 12--16 (2020)

\bibitem{bailey2007relationship}
Bailey, A., Channon, S., Beaumont, J.: The relationship between subjective
  fatigue and cognitive fatigue in advanced multiple sclerosis. Multiple
  Sclerosis Journal  13(1),  73--80 (2007)

\bibitem{belmont2006fatigue}
Belmont, A., Agar, N., Hugeron, C., Gallais, B., Azouvi, P.: Fatigue and
  traumatic brain injury. In: Annales de r{\'e}adaptation et de m{\'e}decine
  physique. vol.~49, pp. 370--374. Elsevier (2006)

\bibitem{bendak2020fatigue}
Bendak, S., Rashid, H.S.: Fatigue in aviation: A systematic review of the
  literature. International Journal of Industrial Ergonomics  76,  102928
  (2020)

\bibitem{braithwaite2013guide}
Braithwaite, J.J., Watson, D.G., Jones, R., Rowe, M.: A guide for analysing
  electrodermal activity (eda) \& skin conductance responses (scrs) for
  psychological experiments. Psychophysiology  49(1),  1017--1034 (2013)

\bibitem{chaudhuri2004fatigue}
Chaudhuri, A., Behan, P.O.: Fatigue in neurological disorders. The lancet
  363(9413),  978--988 (2004)

\bibitem{cifrek2009surface}
Cifrek, M., Medved, V., Tonkovi{\'c}, S., Ostoji{\'c}, S.: Surface emg based
  muscle fatigue evaluation in biomechanics. Clinical biomechanics  24(4),
  327--340 (2009)

\bibitem{dawson2011skin}
Dawson, M.E., Schell, A.M., Courtney, C.G.: The skin conductance response,
  anticipation, and decision-making. Journal of Neuroscience, Psychology, and
  Economics  4(2),  111 (2011)

\bibitem{einthoven1950direction}
Einthoven, W., Fahr, G., De~Waart, A.: On the direction and manifest size of
  the variations of potential in the human heart and on the influence of the
  position of the heart on the form of the electrocardiogram. American heart
  journal  40(2),  163--211 (1950)

\bibitem{garcia2014effects}
Garc{\'\i}a-P{\'e}rez, J.A., P{\'e}rez-Soriano, P., Llana~Belloch, S.,
  Lucas-Cuevas, {\'A}.G., S{\'a}nchez-Zuriaga, D.: Effects of treadmill running
  and fatigue on impact acceleration in distance running. Sports Biomechanics
  13(3),  259--266 (2014)

\bibitem{guastello2015catastrophe}
Guastello, S.J., Reiter, K., Malon, M., Timm, P., Shircel, A., Shaline, J.:
  Catastrophe models for cognitive workload and fatigue in n-back tasks.
  Nonlinear Dynamics, Psychology, and Life Sciences  (2015)

\bibitem{hagell2009towards}
Hagell, P., Brundin, L.: Towards an understanding of fatigue in parkinson
  disease. Journal of Neurology, Neurosurgery \& Psychiatry  80(5),  489--492
  (2009)

\bibitem{hopstaken2015multifaceted}
Hopstaken, J.F., Van Der~Linden, D., Bakker, A.B., Kompier, M.A.: A
  multifaceted investigation of the link between mental fatigue and task
  disengagement. Psychophysiology  52(3),  305--315 (2015)

\bibitem{huang2018detection}
Huang, S., Li, J., Zhang, P., Zhang, W.: Detection of mental fatigue state with
  wearable ecg devices. International journal of medical informatics  119,
  39--46 (2018)

\bibitem{jaiswal2021understanding}
Jaiswal, A., Babu, A.R., Zadeh, M.Z., Makedon, F., Wylie, G.: Understanding
  cognitive fatigue from fmri scans with self-supervised learning. arXiv
  preprint arXiv:2106.15009  (2021)

\bibitem{jap2009using}
Jap, B.T., Lal, S., Fischer, P., Bekiaris, E.: Using eeg spectral components to
  assess algorithms for detecting fatigue. Expert Systems with Applications
  36(2),  2352--2359 (2009)

\bibitem{ji2006probabilistic}
Ji, Q., Lan, P., Looney, C.: A probabilistic framework for modeling and
  real-time monitoring human fatigue. IEEE Transactions on systems, man, and
  cybernetics-Part A: Systems and humans  36(5),  862--875 (2006)

\bibitem{krupp1988fatigue}
Krupp, L.B., Alvarez, L.A., LaRocca, N.G., Scheinberg, L.C.: Fatigue in
  multiple sclerosis. Archives of neurology  45(4),  435--437 (1988)

\bibitem{luo2020assessment}
Luo, H., Lee, P.A., Clay, I., Jaggi, M., De~Luca, V.: Assessment of fatigue
  using wearable sensors: a pilot study. Digital biomarkers  4(1),  59--72
  (2020)

\bibitem{neurokit}
Makowski, D., Pham, T., Lau, Z.J., Brammer, J., Lespinasse, F., Pham, H.,
  Schölzel, C., Chen, S.: Neurokit2: A python toolbox for neurophysiological
  signal processing. Behavior Research Methods  53 (02 2021)

\bibitem{Marcora2009Mental}
Marcora, S.M., Staiano, W., Manning, V.: Mental fatigue impairs physical
  performance in humans. Journal of Applied Physiology  106(3),  857--864
  (2009)

\bibitem{meier2014developmental}
Meier, B., Rothen, N., Walter, S.: Developmental aspects of synaesthesia across
  the adult lifespan. Frontiers in human neuroscience  8,  129 (2014)

\bibitem{muses}
MUSE: Muse s - the next generation of muse s,
  \url{"https://choosemuse.com/muse-s/"}

\bibitem{myles1985sleep}
Myles, W.S.: Sleep deprivation, physical fatigue, and the perception of
  exercise intensity. Medicine \& Science in Sports \& Exercise  (1985)

\bibitem{nelesen2008relationship}
Nelesen, R., Dar, Y., Thomas, K., Dimsdale, J.E.: The relationship between
  fatigue and cardiac functioning. Archives of internal medicine  168(9),
  943--949 (2008)

\bibitem{pan1985real}
Pan, J., Tompkins, W.J.: A real-time qrs detection algorithm. IEEE transactions
  on biomedical engineering pp. 230--236 (1985)

\bibitem{ramesh2020multi}
Ramesh~Babu, A., Zadeh, M.Z., Jaiswal, A., Lueckenhoff, A., Kyrarini, M.,
  Makedon, F.: A multi-modal system to assess cognition in children from their
  physical movements. In: Proceedings of the 2020 International Conference on
  Multimodal Interaction. pp. 6--14 (2020)

\bibitem{richley2013new}
Richley, D.: New training and qualifications in electrocardiography. British
  Journal of Cardiac Nursing  8(1),  38--42 (2013)

\bibitem{Ringnr2008WhatIP}
Ringn{\'e}r, M.: What is principal component analysis? Nature Biotechnology
  26,  303--304 (2008)

\bibitem{rota2014influence}
Rota, S., Morel, B., Saboul, D., Rogowski, I., Hautier, C.: Influence of
  fatigue on upper limb muscle activity and performance in tennis. Journal of
  Electromyography and Kinesiology  24(1),  90--97 (2014)

\bibitem{russell2021predicting}
Russell, B., McDaid, A., Toscano, W., Hume, P.: Predicting fatigue in long
  duration mountain events with a single sensor and deep learning model.
  Sensors  21(16),  5442 (2021)

\bibitem{Sedighi2017data}
{Sedighi Maman}, Z., {Alamdar Yazdi}, M.A., Cavuoto, L.A., Megahed, F.M.: A
  data-driven approach to modeling physical fatigue in the workplace using
  wearable sensors. Applied Ergonomics  65,  515--529 (2017)

\bibitem{Xu2018How}
Xu, R., Zhang, C., He, F., Zhao, X., Qi, H., Zhou, P., Zhang, L., Ming, D.: How
  physical activities affect mental fatigue based on eeg energy, connectivity,
  and complexity. Frontiers in Neurology  9 (2018)

\bibitem{zadeh2020towards}
Zadeh, M.Z., Babu, A.R., Lim, J.B., Kyrarini, M., Wylie, G., Makedon, F.:
  Towards cognitive fatigue detection from functional magnetic resonance
  imaging data. In: Proceedings of the 13th ACM International Conference on
  PErvasive Technologies Related to Assistive Environments. pp. 1--2 (2020)

\end{thebibliography}

\end{document}